\documentclass[fleqn,usenatbib]{mnras}

\usepackage{newtxtext,newtxmath}

\usepackage[T1]{fontenc}
\usepackage{ae,aecompl}

\usepackage{graphicx}	
\usepackage{amsmath}	
\usepackage{amssymb}	
\usepackage[usenames,dvipsnames]{xcolor}

\newcommand{\kms}{\mbox{$\mathrm{km\,s^{-1}}$}}
\newcommand{\MSUN}{\mbox{$\mathrm{M_{\odot}}$}}
\newcommand{\MJUP}{\mbox{$\mathrm{M_J}$}}
\newcommand{\RSUN}{\mbox{$\mathrm{R_{\odot}}$}}
\newcommand{\TEFF}{\mbox{$\mathrm{T_{eff}}$}}

\title[Two ultrashort eclipsing binaries from K2]{Two white dwarfs in ultrashort binaries with detached, eclipsing, likely substellar companions detected by {\em K2}}

\author[S. G. Parsons et al.]{S.~G.~Parsons$^{1}$\thanks{s.g.parsons@sheffield.ac.uk},
J.~J.~Hermes$^{2}$\thanks{Hubble Fellow},
T.~R.~Marsh$^{3}$,
B.~T.~G{\"a}nsicke$^{3}$,
P.-E.~Tremblay$^{3}$,
\newauthor
S.~P.~Littlefair$^{1}$,
D.~I.~Sahman$^{1}$,
R.~P.~Ashley$^{3}$,
M.~Green$^{3}$,
S.~Rattanasoon$^{1,4}$,
\newauthor
V.~S.~Dhillon$^{1,5}$,
M.~R.~Burleigh$^{6}$,
S.~L.~Casewell$^{6}$,
D.~A.~H.~Buckley$^{7}$,
I.~P.~Braker$^{6}$,
\newauthor
P.~Irawati$^{4}$,
E.~Dennihy$^{2}$,
P.~Rodr\'iguez-Gil$^{5,8}$,
D.~E.~Winget$^{9,10}$,
K.~I.~Winget$^{9,10}$,
\newauthor
Keaton~J.~Bell$^{9,10}$, and
Mukremin~Kilic$^{11}$
\\
$^{1}$ Department of Physics and Astronomy, University of Sheffield,
Sheffield, S3 7RH, UK\\
$^{2}$ Department of Physics and Astronomy, University of North Carolina, Chapel Hill, NC 27599-3255, USA\\
$^{3}$ Department of Physics, University of Warwick, Coventry CV4 7AL, UK\\
$^{4}$ National Astronomical Research Institute of Thailand, 191 Siriphanich
Bldg., Huay Kaew Road, Chiang Mai 50200, Thailand\\
$^{5}$ Instituto de Astrof{\'i}sica de Canarias, V{\'i}a Lactea s/n, La
Laguna, E-38205 Tenerife, Spain\\
$^{6}$ Department of Physics and Astronomy, University of Leicester, Leicester
LE1 7RH, UK\\
$^{7}$ South African Astronomical Observatory, PO Box 9, Observatory, 7935,
Cape Town, South Africa\\
$^{8}$ Universidad de La Laguna, Departamento de Astrof\'isica, E-38206 La Laguna, Tenerife, Spain\\
$^{9}$ Department of Astronomy, University of Texas at Austin, Austin, TX 78712, USA\\
$^{10}$ McDonald Observatory, Fort Davis, TX 79734, USA\\
$^{11}$ Homer L. Dodge Department of Physics and Astronomy, University of Oklahoma, Norman, OK 73019, USA\\
}

\date{Accepted 2017 June 22. Received 2017 June 16; in original form 2017 April 28}

\pubyear{2017}

\begin{document}
\label{firstpage}
\pagerange{\pageref{firstpage}--\pageref{lastpage}}
\maketitle

\begin{abstract}
Using data from the extended {\em Kepler} mission in {\em K2} Campaign 10 we identify two eclipsing binaries containing white dwarfs with cool companions that have extremely short orbital periods of only 71.2 min (SDSS\,J1205$-$0242, a.k.a. EPIC\,201283111) and 72.5 min (SDSS\,J1231+0041, a.k.a. EPIC\,248368963). Despite their short periods, both systems are detached with small, low-mass companions, in one case a brown dwarf, and the other case either a brown dwarf or a low-mass star. We present follow-up photometry and spectroscopy of both binaries, as well as phase-resolved spectroscopy of the brighter system, and use these data to place preliminary estimates on the physical and binary parameters. SDSS\,J1205$-$0242 is composed of a $0.39\pm0.02${\MSUN} helium-core white dwarf which is totally eclipsed by a $0.049\pm0.006${\MSUN} ($51\pm6$\MJUP) brown dwarf companion, while SDSS\,J1231+0041 is composed of a $0.56\pm0.07${\MSUN} white dwarf which is partially eclipsed by a companion of mass $\lesssim 0.095${\MSUN}. In the case of SDSS\,J1205$-$0242 we look at the combined constraints from common-envelope evolution and brown dwarf models; the system is compatible with similar constraints from other post common-envelope binaries given the current parameter uncertainties, but has potential for future refinement.
\end{abstract}




\section{Introduction}

\begin{table*}
 \centering
  \caption{Journal of observations. The eclipse of the white dwarf occurs at
    Phase 1, 2, etc.}
  \label{tab:obs}
  \begin{tabular}{@{}lccccccc@{}}
  \hline
  Date at     & Telescope/ &Filter &Start   & Orbital     &Exposure   & Number of & Conditions               \\
  start of run& Instrument &       &(UT)    & phase       &time (s)   & exposures & (Transparency, seeing)   \\
  \hline
  \multicolumn{2}{l}{\bf SDSS\,J1205$-$0242:}\\
  2016/07/13  & {\em K2} Campaign 10 & {\em Kepler} & 02:09 & ...         & 1765.3 & 2656 & Space-based \\
  2017/01/03  & McDonald/ProEM       & $r'$         & 09:46 & $0.84-3.50$ & 30.0   & 380  & Clear, $\sim$1.7 arcsec \\
  2017/01/19  & TNT/ULTRASPEC        & $i'$         & 19:43 & $0.68-2.31$ & 14.5   & 481  & Clear, $\sim$1.5 arcsec \\
  2017/01/19  & TNT/ULTRASPEC        & {\em kg5}    & 20:59 & $0.64-1.19$ & 8.5    & 277  & Clear, $\sim$1.0 arcsec \\
  2017/01/22  & TNT/ULTRASPEC        & {\em kg5}    & 20:15 & $0.78-1.27$ & 8.5    & 244  & Clear, $\sim$1.2 arcsec \\
  2017/01/25  & TNT/ULTRASPEC        & {\em kg5}    & 19:31 & $0.81-2.14$ & 10.0   & 575  & Clear, $\sim$1.5 arcsec \\
  2017/01/26  & SOAR/Goodman         & -            & 07:36 & $0.99-2.21$ & 300.0  & 17   & Clear, $\sim$1.4 arcsec \\
  2017/01/29  & TNT/ULTRASPEC        & {\em clear}  & 17:45 & $0.80-1.34$ & 10.0   & 614  & Some clouds, $\sim$2.0 arcsec \\
  2017/02/18  & TNT/ULTRASPEC        & $z'$         & 20:25 & $0.76-1.18$ & 12.8   & 142  & Clear, $\sim$2.0 arcsec \\
  2017/02/19  & TNT/ULTRASPEC        & $z'$         & 17:41 & $0.68-1.05$ & 12.8   & 121  & Clear, $\sim$2.0 arcsec \\
  2017/02/22  & TNT/ULTRASPEC        & $z'$         & 16:45 & $0.53-1.09$ & 15.0   & 163  & Clear, $\sim$2.0 arcsec \\
  2017/03/04  & SALT/SALTICAM        & $i'$         & 22:00 & $0.22-0.73$ & 10.0   & 216  & Clear, $\sim$1.4 arcsec \\
  2017/03/27  & GTC/OSIRIS           & -            & 23:27 & $0.44-1.87$ & 240.5  & 25   & Clear, $\sim$1.2 arcsec \\
  \multicolumn{2}{l}{\bf SDSS\,J1231$+$0041:}\\
  2016/07/13  & {\em K2} Campaign 10 & {\em Kepler} & 02:10 & ...         & 1765.3 & 2649 & Space-based \\
  2017/01/20  & TNT/ULTRASPEC        & {\em kg5}    & 21:15 & $0.22-1.69$ & 20.0   & 329  & Clear, $\sim$1.0 arcsec \\
  2017/01/22  & TNT/ULTRASPEC        & {\em kg5}    & 20:53 & $0.56-2.40$ & 20.0   & 410  & Clear, $\sim$1.2 arcsec \\
  2017/02/21  & TNT/ULTRASPEC        & {\em kg5}    & 17:18 & $0.40-2.20$ & 15.0   & 529  & Clear, $\sim$2.0 arcsec \\
  2017/02/22  & TNT/ULTRASPEC        & {\em kg5}    & 17:45 & $0.53-2.14$ & 15.0   & 548  & Clear, $\sim$2.0 arcsec \\
  2017/03/01  & SOAR/Goodman         & -            & 05:33 & $0.54-2.42$ & 600.0  & 13   & Clear, $\sim$1.5 arcsec \\
  \hline
\end{tabular}
\end{table*}

Roughly 75 per cent of main-sequence binaries are born wide enough that they evolve essentially as single stars \citep{willems04}. However, for the remaining 25 per cent, the expansion of the more massive star at the end of its main-sequence lifetime causes the two stars to interact, often initiating a common-envelope event. The frictional forces experienced by the stars during the common-envelope phase result in a dramatic reduction in the separation of the two stars, down to as low as a few solar radii. Angular momentum loss via magnetic braking and gravitational radiation drives the resulting post-common-envelope binary (PCEB) to shorter periods, eventually creating a cataclysmic variable system. 

Large-scale surveys have led to an explosion in the number of known PCEBs \citep{silvestri2006, rebassa-mansergas2010, rebassa16}, with more than 100 systems having measured periods \citep[e.g.][]{nebot11,parsons15}. The vast majority of these systems have M-dwarf companions, with the spectral type distribution of the secondaries peaking near M3--M4, in good agreement with the peak in the initial mass function of single stars \citep{chabrier2003}. Just six PCEBs are known to be composed of white dwarfs with brown-dwarf companions \citep{dobbie05,burleigh06,casewell12,steele13,littlefair14,farihi17}.

PCEBs with brown-dwarf companions are difficult to identify from optical data alone. However, infrared surveys have demonstrated that these systems are intrinsically rare, with only 0.4$-$2 per cent of white dwarfs having a sub-stellar companion \citep{farihi2005, girven11, steele11,debes11}, including both wide binaries that never interacted (e.g. \citealt{becklin+zuckerman88, steele2009, luhman2011}) and PCEBs. The small number of white dwarfs with brown dwarf companions reflects the rarity of sub-stellar objects both in the field \citep{kirkpatrick2012} and in binaries \citep{grether+lineweaver2006}.

The short orbital periods of PCEBs provide for many deeply eclipsing binaries, which offer a unique opportunity to directly probe the structures of both components by allowing for model-independent, high-precision mass and radius measurements (e.g., \citealt{parsons12b}). This is especially useful for uncommon objects. For example, there are very few known eclipsing binaries composed of at least one brown dwarf.

There is only one double-lined, eclipsing brown dwarf binary known to date, 2MASS\,J05352184$-$0546085 \citep{stassun06}, although another has tentatively been identified \citep{david2016}. Both of these systems are young ($<$10\,Myr), however, which will affect their radii, as brown dwarfs contract throughout their lifetime. The remainder of brown dwarfs with direct measurements of their radii are in systems where they are highly irradiated. For example, Kelt-1b is a 27\,{\MJUP} brown dwarf in a 29-hr orbit around an F star \citep{siverd2012}; it is known to be highly inflated, at the 10$\sigma$ level compared to models. However, Wasp-30b \citep{anderson11} is a 61{\MJUP} brown dwarf orbiting an F8 star every 4.16\,d and has a radius that agrees with model predictions. SDSS\,J141126.20+200911.1, the only known brown dwarf to be eclipsing in a detached PCEB, also has a radius which is consistent with model predictions \citep{littlefair14}.

There is thus considerable value in finding more eclipsing PCEBs containing a brown dwarf. As part of a search for transits and variability in white dwarfs observed during {\em K2} Campaign 10, we have discovered two new eclipsing PCEBs composed of a white dwarf and a likely brown dwarf companion. {\em K2} \citep{howell14} is an extension of the {\em Kepler} planet-hunting mission \citep{borucki10} in which a number of fields along the ecliptic are continuously observed with high photometric precision over a period of approximately 75 days, hence it is ideal for detecting eclipsing PCEBs. We report here follow-up photometry and spectroscopy for these two new eclipsing systems, and furthermore detail and constrain their binary and stellar parameters.

\section{Observations and their reduction}

A full journal of observations is given in Table~\ref{tab:obs}. 

\subsection{Target selection}

We have proposed multiple Guest Observer programs to search for transits and variability from hundreds of known and candidate white dwarfs in every campaign of the {\em K2} mission.
As part of an analysis of targets observed with long-cadence (29.4-min) exposures during {\em K2} Campaign 10, we flagged two spectroscopically confirmed white dwarfs with high-amplitude, short-period variability. The first, SDSS\,J120515.80$-$024222.6 (a.k.a. EPIC\,201283111\footnote{Proposed by {\em K2} Guest Observer programs led by PIs Kilic (GO10006), Hermes (GO10018), and Burleigh (GO10019).}, hereafter SDSS\,J1205$-$0242), showed variability near 71.2\,min, very near the Nyquist frequency of our dataset. The other, SDSS\,J123127.14+004132.9 (a.k.a. EPIC\,248368963\footnote{Targeted by the program led by PI Kilic (GO10006).}, henceforth SDSS\,J1231+0041), showed variability at a similarly short period of 72.5\,min.

\subsection{{\em K2} photometry}

We examined preliminary extractions of our known and candidate white dwarfs using light curves produced by the {\em Kepler} Guest Observer (GO) office \citep{vancleve16}, available through the Barbara A. Mikulski Archive for Space Telescopes (MAST). The {\em Kepler} bandpass covers roughly 4000$-$9000\,\AA. Each {\em K2} long-cadence observation represents a co-add of 270$\times$6.02 s exposures.

We improved our extraction of SDSS\,J1205$-$0242 ($K_p$=18.8\,mag) by downloading the processed target pixel file from MAST and using the {\sc PyKE} software tools \citep{still12}. Using a large (17-px) fixed aperture, we extracted the light curve, fit out a quadratic function to 3-day windows, and corrected for {\em K2} motion artifacts using the {\sc kepsff} software package \citep{vanderburg14}. Subsequently, we clipped by-hand any highly discrepant points. All data obtained in {\em K2} Campaign 10 suffer from a large gap caused by the failure of a CCD module onboard the spacecraft roughly 7\,d into the campaign, which powered off the photometer for roughly 14\,d. Still, our final 69.12-d light curve of SDSS\,J1205$-$0242 has 2656 points and a duty cycle exceeding 78 per cent.

For SDSS\,J1231+0041 ($K_p$=20.0\,mag), we saw little improvement with our custom {\sc PyKE} extraction, and used the light curve produced by the GO office for our final dataset, which was extracted with a 2-px aperture. After clipping, the final 69.12-day light curve of SDSS\,J1231+0041 has 2644 points.

\subsection{McDonald+ProEM photometry}

We obtained the first follow-up data of SDSS\,J1205$-$0242 on 2017 January 3 using the ProEM frame-transfer camera mounted at the Cassegrain focus of the 2.1-m Otto Struve telescope at McDonald Observatory in West Texas. The data were collected through an $r'$ filter. We performed differential, circular aperture photometry by extracting the target and a nearby comparison star using the {\sc IRAF} task {\sc CCD\_HSP} \citep{kanaan02}, and applied a barycentric correction using the {\sc WQED} software package \citep{thompson13}.

\subsection{TNT+ULTRASPEC photometry}

We observed both our targets with the high-speed frame-transfer EMCCD camera ULTRASPEC \citep{dhillon14} mounted on the 2.4-m Thai National Telescope (TNT) on Doi Inthanon, Thailand in January and February 2017. Our observations were made using the $i'$ band filter, a broad $u'$+$g'$+$r'$ filter known as {\em kg5} (as described in \citealt{dhillon14}, see also the appendix of \citealt{hardy17}), as well as the $z'$ band and ``clear'' (fused silica) filters. Exposure times were adjusted depending upon the conditions. The dead time between each exposure is 15\,ms. All of these data were reduced using the ULTRACAM pipeline software. The source flux was determined with aperture photometry using a variable aperture scaled according to the full width at half maximum. Variations in observing conditions were accounted for by determining the flux relative to a comparison star in the field of view.

\subsection{SOAR+Goodman spectroscopy}

To better constrain the atmospheric parameters of the primary white dwarfs in both systems, we obtained low-resolution spectra of the upper Balmer series using the high-throughput Goodman spectrograph \citep{clemens04} on the 4.1-m SOAR telescope at Cerro Pach\'{o}n in Chile. We used a 930 line mm$^{-1}$ grating, and our setup covers roughly $3600-5200${\AA} at a resolution of roughly 4{\AA}, set by the seeing.

Using a 1.69\arcsec\ slit, we obtained 18$\times$300\,s exposures of SDSS\,J1205$-$0242 on 2017 January 26. The data were optimally extracted \citep{marsh89} using the {\sc pamela} package within {\sc starlink} and flux calibrated using the standard Feige\,67. The final summed spectrum has a signal-to-noise (S/N) of 65 per resolution element in the continuum around 4600{\AA}. We obtained 13$\times$600\,s exposures of SDSS\,J1231+0041 on 2017~March~1 using a 3.0\arcsec\ slit. The optimally extracted spectra were flux calibrated against LTT\,2415 and have a summed S/N of 24 per resolution element around 4600{\AA}. SOAR spectroscopy of both targets were obtained at minimal airmass.

\subsection{SALT+SALTICAM photometry}

We obtained time-series photometry of SDSS\,J1205$-$0242 using the high-speed camera SALTICAM \citep{2006MNRAS.372..151O} mounted on the 10-m Southern African Large Telescope (SALT) on 2017 March 4. We used SALTICAM in the frame-transfer mode, whereby the moving mask occults half the CCD (the storage array), and we took 10\,s exposures with 4$\times$4 binning, yielding a plate scale of 0.56 arcsec/pixel. All SALTICAM observations had essentially zero deadtime ($<$6\,ms) between frames.

\subsection{GTC+OSIRIS spectroscopy}

We observed SDSS\,J1205$-$0242 with the Optical System for Imaging and low-Intermediate-Resolution Integrated Spectroscopy (OSIRIS) on the 10.4-m Gran Telescopio Canarias (GTC) on La Palma. We used the R2500R grism with a 0.6\arcsec\ slit centred on the H$\alpha$ line, giving a resolution of R$\simeq$2500. We used exposure times of 240\,s and recorded a total of 25 spectra of SDSS\,J1205$-$0242, as well as one spectrum of the spectrophotometric standard star Hiltner\,600.

The data were optimally extracted using {\sc pamela}. An arc spectrum was used to wavelength calibrate the data. In total 34 lines (mostly neon) were fitted with a sixth-order polynomial giving an rms of 0.01{\AA}. We then applied additional pixel shifts to each exposure (0.3 pixels maximum) based on the positions of three skylines (6300{\AA}, 6863{\AA} and 7276{\AA}) to correct for instrument flexure. Finally, the instrumental response was removed using the spectrum of the standard star.

\section{A 71.2-min Binary: SDSS J1205-0242}
\label{sec:1205}

SDSS\,J1205$-$0242 ($g=18.5$\,mag) was classified as a white dwarf based on a serendipitous spectrum from the fourth data release of the Sloan Digital Sky Survey (SDSS) by \citet{eisenstein06}. The SDSS spectrum shows no sign of a companion and no obvious red excess, but an automated fit to the spectrum yields a mass of $0.39\pm0.03${\MSUN} \citep{kleinman13}, which is extremely low for a white dwarf, implying a binary origin \citep{marsh95,rebassa-mansergas2011}.

\begin{figure}
  \begin{center}
    \includegraphics[angle=-90,width=\columnwidth]{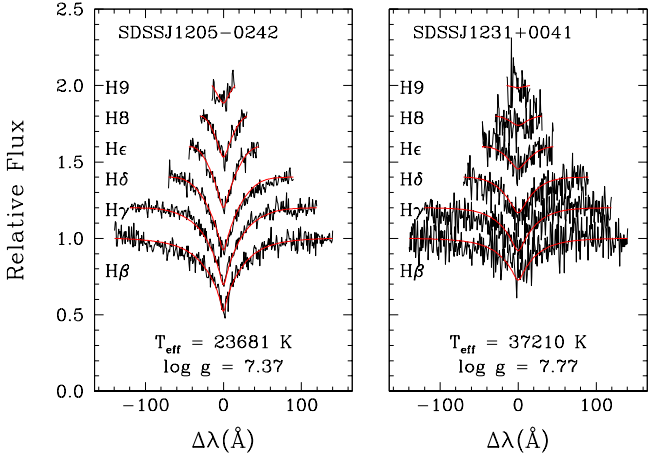}
    \vspace{-4pt}
    \caption{The averaged spectrum of SDSS\,J1205$-$0242 (left) and SDSS\,J1231+0041 (right) obtained with the Goodman spectrograph on the 4.1-m SOAR telescope. Our best fits to the Balmer lines (shown in red) yield updated atmospheric parameters for the primary white dwarfs in each system.} 
  \label{fig:soar_spec}
  \end{center}
\end{figure}

To confirm if this white dwarf genuinely has a low mass we analyzed our follow-up, higher-S/N spectrum from SOAR using the fitting procedures and pure-hydrogen, one-dimensional model atmospheres for white dwarfs described in \citet{tremblay11}. The fit to our SOAR spectrum is consistent with the SDSS fit, giving a temperature of $\TEFF=23680\pm430$\,K and surface gravity of $\log{g}=7.374\pm0.057$, implying a mass of $0.39\pm0.02${\MSUN} and cooling age of 50\,Myr calculated using the helium-core white dwarf models of \citet{panei07}. The Balmer lines and best fit to SDSS\,J1205$-$0242 are shown in the left panel of Fig.~\ref{fig:soar_spec}. 

\begin{figure}
  \begin{center}
    \includegraphics[width=\columnwidth]{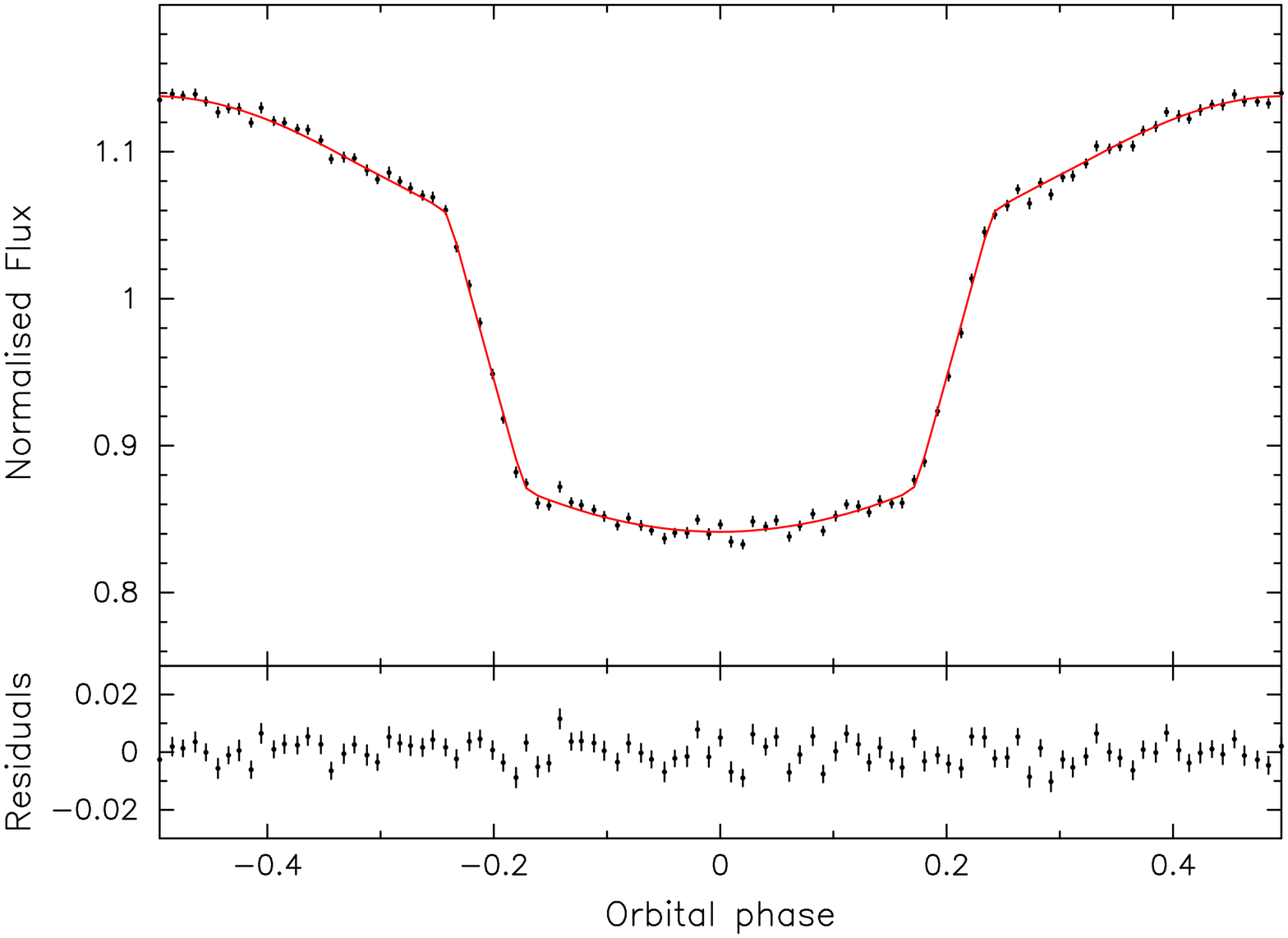}
    \includegraphics[width=\columnwidth]{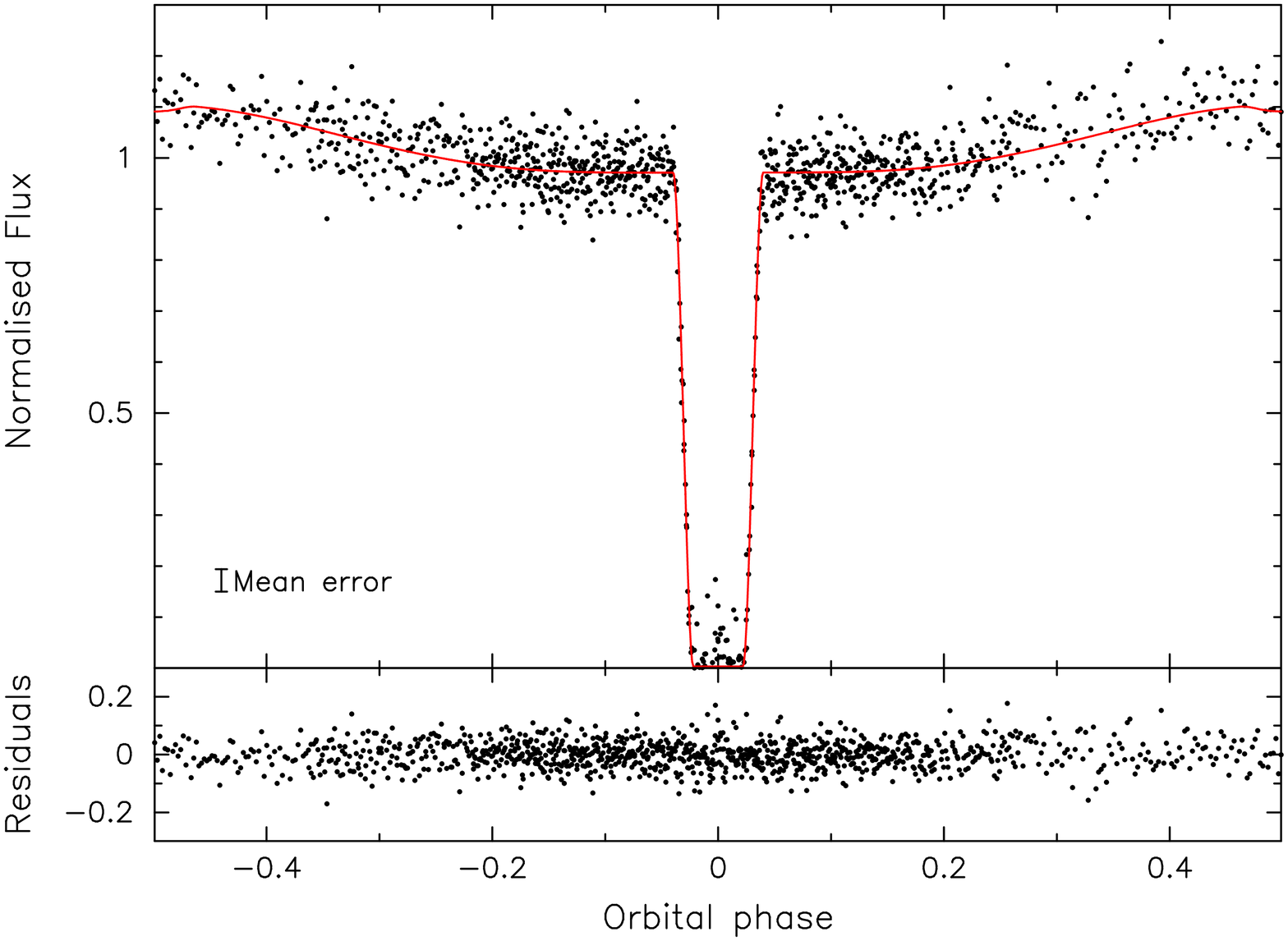}
    \vspace{-12pt}
    \caption{{\bf Top:} Phase-folded, binned {\em K2} light curve of SDSS\,J1205$-$0242. It is strongly smeared, since each long-cadence exposure lasts 29.4\,min of the 71.2-min orbit. Still, a sinusoidal variation and eclipse stand out. {\bf Bottom:} Phase-folded, ULTRASPEC {\em kg5} light curve of SDSS\,J1205$-$0242 with higher time sampling. The deep eclipse of the white dwarf is clear, as is the out-of-eclipse reflection effect. There is no detection of the companion during the eclipse. Over-plotted in red is the best-fit model light curve (phase smearing is accounted for in the {\em K2} plot at top).} 
  \label{fig:sdss1205_lc}
  \end{center}
\end{figure}

The {\em K2} light curve (Fig.~\ref{fig:sdss1205_lc}) shows strong variations on a period of 0.04946539(9)\,days (71.2\,min). The {\em K2} data were obtained in long-cadence mode (29.4-min exposures), so any sharp features in the light curve are significantly smeared out. Nevertheless, a clear reflection effect is seen, along with steep eclipse features, implying that the system is likely to be fully eclipsing.

We display in the bottom panel of Fig.~\ref{fig:sdss1205_lc} our high-speed, follow-up ULTRASPEC light curve of SDSS\,J1205$-$0242, which shows a deep eclipse of the white dwarf, lasting just 5\,min from the first to fourth contact points. The reflection effect is also evident out of eclipse. We recorded a total of seven eclipses, although due to the low S/N of the three $z'$ band eclipses we excluded these from our ephemeris calculations. The four remaining eclipse times are listed in Table~\ref{tab:etimes}. We fitted each of these eclipses with a code specifically designed for binaries containing at least one white dwarf \citep{copperwheat10} in order to determine the mid-eclipse times. From these we determined the ephemeris for the system to be
\begin{eqnarray}
\mathrm{MJD(BTDB)} = 57768.039311(3) + 0.049465250(6)E,
\end{eqnarray}
where $E$ is the orbital phase, with $E=0$ corresponding to the centre of the white dwarf eclipse.

\begin{figure*}
  \begin{center}
    \includegraphics[width=0.75\textwidth]{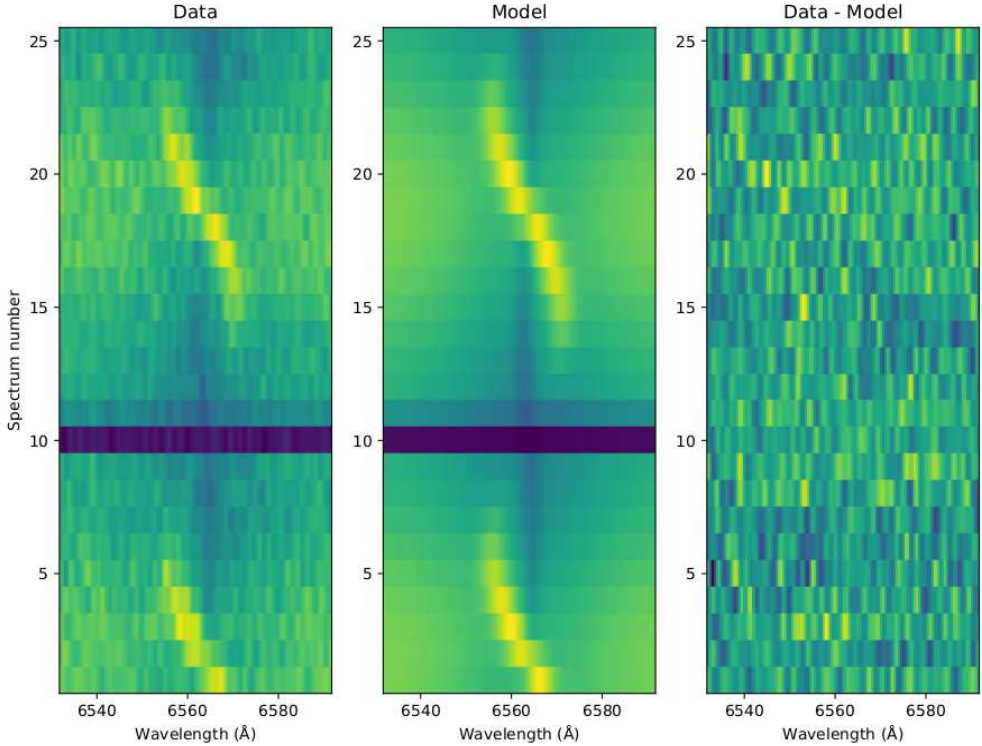}
    \vspace{-8pt}
    \caption{Trailed spectra of the H$\alpha$ line of SDSS\,J1205$-$0242 with time running upwards. The left-hand panel shows our GTC/OSIRIS data (the eclipse of the white dwarf occurs during spectrum 10). The centre panel shows our best fit model to the line, including both the absorption from the white dwarf and the emission from its companion. The right-hand panel shows the residuals of the fit.} 
  \label{fig:trail}
  \end{center}
\end{figure*}

\begin{table}
 \centering
  \caption{Mid-eclipse times.}
  \label{tab:etimes}
  \begin{tabular}{@{}lcc@{}}
  \hline
  Cycle & MJD(BTDB) & Source \\
  \hline
  \multicolumn{3}{l}{\bf SDSS\,J1205$-$0242:}\\
  $-$3061& 57616.626169(18) & {\em K2} Campaign 10 \\
  $-$234 & 57756.4644612(78) & McDonald $r'$ band \\
  97   & 57772.837463(30) & ULTRASPEC $i'$ band \\
  118  & 57773.8762091(87) & ULTRASPEC {\em kg5} band \\
  158  & 57775.8548125(63) & ULTRASPEC {\em kg5} band \\
  218  & 57778.8227308(78) & ULTRASPEC {\em kg5} band \\
  301  & 57782.9283510(78) & ULTRASPEC clear \\
  \multicolumn{3}{l}{\bf SDSS\,J1231+0041:}\\
  $-$2575& 57616.774090(86) & {\em K2} Campaign 10 \\
  546  & 57773.928207(40) & ULTRASPEC {\em kg5} band \\
  585  & 57775.892041(30) & ULTRASPEC {\em kg5} band \\
  1178 & 57805.751893(35) & ULTRASPEC {\em kg5} band \\
  1198 & 57806.758914(31) & ULTRASPEC {\em kg5} band \\
  \hline
\end{tabular}
\end{table}

To further improve the physical constraints of each component of the binary, we obtained GTC+OSIRIS time-series spectroscopy of the H$\alpha$ line (left-hand panel of Fig.~\ref{fig:trail}). The spectroscopy shows both a clear absorption component from the white dwarf as well as an emission component moving in anti-phase that is strongest around Phase 0.5 and disappears near the eclipse, the classic signature of irradiation-induced emission lines from the inner face of the companion to the white dwarf.

We fitted the H$\alpha$ line with the following components: (1) A second-order polynomial representing the continuum of the white dwarf, which is scaled according to the light-curve model during phases affected by the eclipse. (2) A first-order polynomial representing the irradiation. Its level is modulated as $(1-\cos{\phi})/2$, where $\phi$ is the orbital phase. (3) Three Gaussian absorption components for the white dwarf, that change position according to $\gamma_1 + K_1\sin{(2\pi\phi)}$. (4) Two Gaussian emission components from the companion star, with strengths modulated in the same way as the irradiation component and that change position according to $\gamma_2 + K_{\mathrm{em}}\sin{(2\pi\phi)}$. We also take into account the smearing of the lines caused by the finite exposure times, and the best model was found using the Levenburg-Marquardt minimisation method.

Our best-fit model is shown in the centre panel of Fig.~\ref{fig:trail}, with the residuals of the fit plotted in the right-hand panel. Our best fit parameters were $\gamma_1=38.5\pm3.5$\,\kms, $K_1=48.3\pm5.1$\,\kms, $\gamma_2=31.9\pm2.6$\,\kms, and $K_{\mathrm{em}}=345.0\pm4.4$\,\kms. The offset between the radial-velocity amplitudes of the two stars $\gamma_1-\gamma_2=6.6\pm4.3$\,{\kms} is effectively the gravitational redshift of the white dwarf. For a 0.39{\MSUN} white dwarf the expected gravitational redshift is 11.2\,{\kms} \citep{panei07}. Correcting this value for the redshift of companion star, the difference in transverse Doppler shifts, and the potential at the companion owing to the white dwarf reduces this to 10.0\,\kms, within $1\sigma$ of the measured value. This provides an external consistency check on the spectroscopically determined mass.

The implied mass ratio of the binary is $q=M_2/M_1=K_1/K_{\mathrm{em}}=0.14$. Assuming a white dwarf mass of 0.39{\MSUN} gives a companion mass of 0.055{\MSUN} or 57{\MJUP}. However, since this emission line originates only from the heated hemisphere of the companion, $K_{\mathrm{em}}$ does not represent its true centre-of-mass velocity, but rather a lower limit on $K_2$, the true radial velocity semi-amplitude of the companion. Therefore, 0.055{\MSUN} represents an upper limit on the mass of the companion; the companion in SDSS\,J1205$-$0242 is therefore definitely substellar. 

The radial velocity amplitude of the companion's centre-of-mass, $K_2$, is related to $K_{\mathrm{em}}$ via the formula
\begin{eqnarray}
K_2 = \frac{K_\mathrm{em}}{1-f(1+q)\frac{R_2}{a}}, \label{eqn:kcorr}
\end{eqnarray}
where $R_2/a$ is the radius of the brown dwarf scaled by the orbital separation ($a$) and $f$ is a constant between 0 and 1 which depends upon the location of the centre of light \citep{parsons12}. We assume a value of $f=0.5$, which roughly corresponds to optically thick emission from the inner hemisphere, which is what has been found for H$\alpha$ emission in similar systems \citep{parsons12b}. This can be combined with the light curve fit to determine a more accurate value of $K_2$.

The combination of the eclipse light curve and the radial velocity information enables us to place constraints on the stellar and binary parameters. When fitting data of the white dwarf eclipse alone there is a degeneracy between the inclination and both stellar radii (scaled by the orbital separation). However, we can establish the relationship between the masses and radii as a function of inclination. To do this we fitted the phase-folded light curve with a binary model \citep[see][for details of the code]{copperwheat10}. We fixed the mass ratio to 0.14 (maximum value from the spectroscopy) and the temperature of the white dwarf was fixed at 24\,000\,K. We used Claret 4-parameter limb-darkening coefficients for a 25000\,K $\log g =7.5$ white dwarf \citep{gianninas13} in the {\em kg5} filter\footnote{Limb-darkening parameters in the $kg5$ filter kindly provided by Alex Gianninas.}. The limb-darkening parameters of the brown dwarf have a negligible impact on the eclipse profile and so were fixed at the linear value for a 2400\,K $\log g =5.0$ star in the SDSS $r$ band \citep{claret12}. The brown dwarf temperature was also fixed at 2400\,K; again this makes no difference to the eclipse fit, since it is undetected during totality. We then varied the inclination from 90 to 84 degrees in steps of 1 degree and allowed the scaled radii, $R_1/a$ and $R_2/a$, to vary. At each inclination we then used the value of $R_2/a$ to compute $K_2$ via Eq.~\ref{eqn:kcorr} and combined this with $K_1$ and the inclination to determine the two masses, as well as $a$ and hence the two radii.

\begin{figure}
  \begin{center}
    \includegraphics[width=0.995\columnwidth]{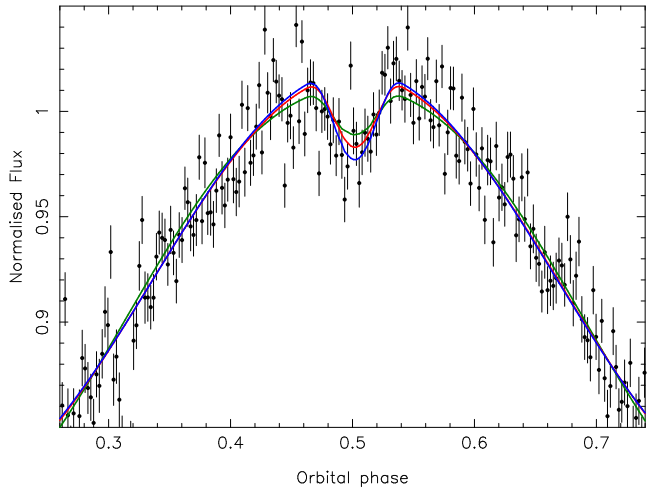}
    \vspace{-12pt}
    \caption{SALTICAM $i'$ band light curve of the transit of the white dwarf in front of the heated face of the brown dwarf (i.e. the secondary eclipse). Also shown are models with inclinations of $90^{\circ}$ (blue), $87^{\circ}$ (red) and $85^{\circ}$ (green). Inclinations lower than $87^{\circ}$ predict a secondary eclipse that is too shallow.}
  \label{fig:sececl}
  \end{center}
\end{figure}

\begin{figure}
  \begin{center}
    \includegraphics[width=0.995\columnwidth]{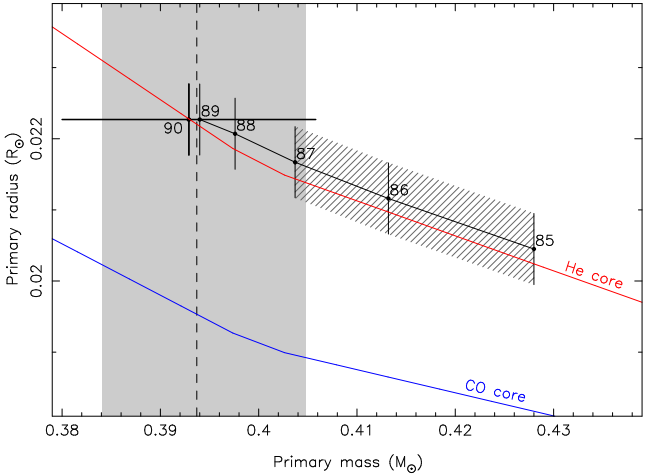}
    \vspace{-12pt}
    \caption{Constraints on the mass and radius of the white dwarf in SDSS\,J1205$-$0242 based on our radial velocity and eclipse fitting. The black line shows how the mass and radius varies as a function of inclination. The red line shows the theoretical mass-radius relationship for a $24\,000$\,K helium-core white dwarf with a canonically thick surface hydrogen layer \citep{panei07}. The blue line shows the same theoretical mass-radius relationship but for a carbon-oxygen core white dwarf \citep{fontaine01}, also with a canonically thick surface hydrogen layer. The vertical dashed line marks the white dwarf mass as determined from the Balmer-line fits to our SOAR spectrum, with the shaded area showing the the 1$\sigma$ uncertainties on this fit. The hatched region shows the inclinations excluded by the secondary eclipse data. The uncertainty on the white dwarf's mass from the radial velocity data is shown on the $90^{\circ}$ model.}
  \label{fig:sdss1205_wdmr}
  \end{center}
\end{figure}

\begin{figure}
  \begin{center}
    \includegraphics[width=\columnwidth]{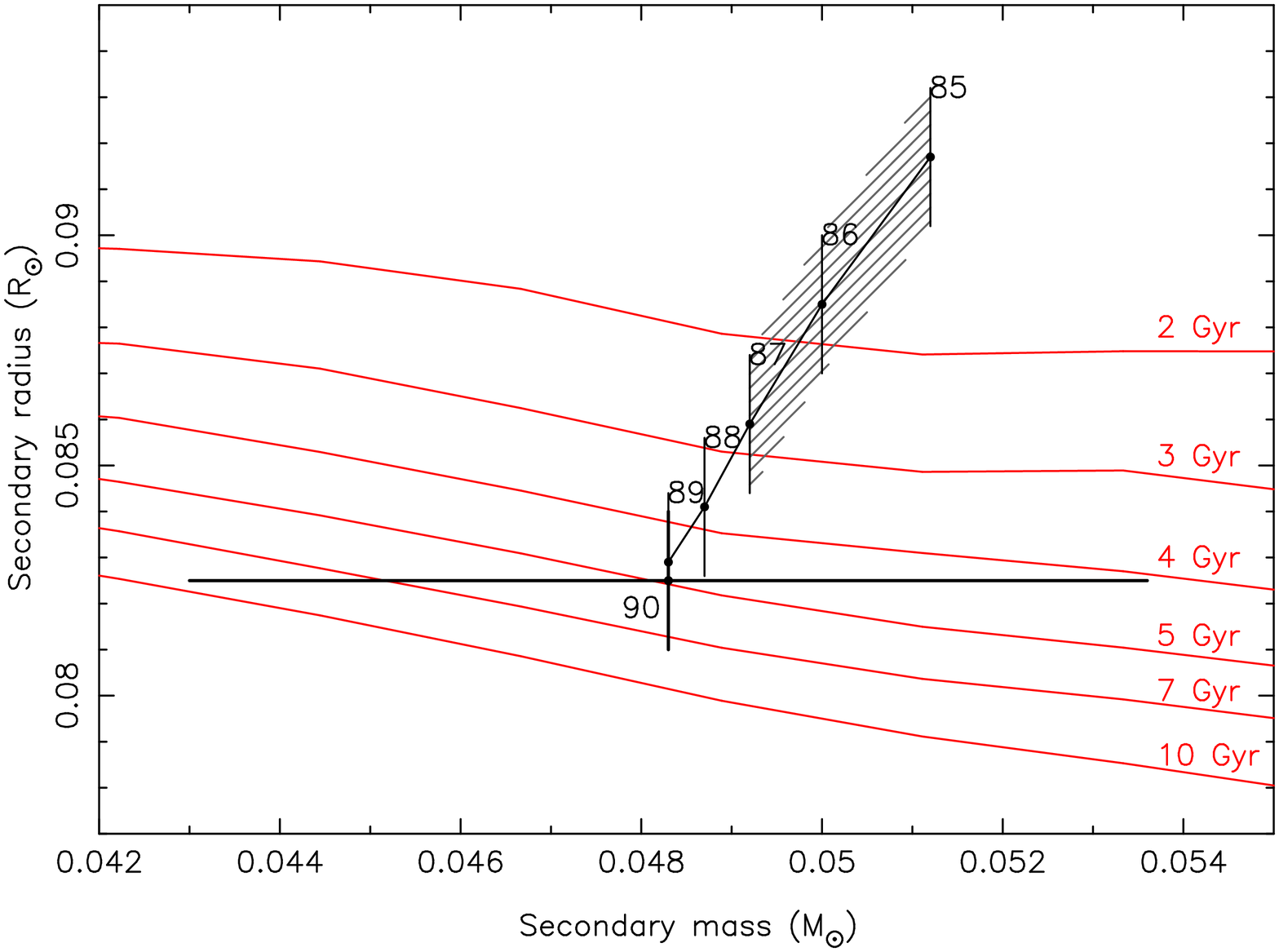}
    \vspace{-12pt}
    \caption{Constraints on the mass and radius of the brown dwarf in SDSS\,J1205$-$0242. The black line shows how the mass and radius varies as a function of inclination given our observational constraints. Note that the radius measurements correspond to the volume-averaged radius of the brown dwarf. Inclinations less than 85$^{\circ}$ are ruled out, as the brown dwarf would fill its Roche lobe. The hatched region shows the inclinations excluded by the secondary eclipse data. Also shown in red are theoretical mass-radius relationships for solar metallicity brown dwarfs of different ages \citep{baraffe03}. The uncertainty on the brown dwarf's mass from the radial velocity data (which dominates over the inclination uncertainty) is shown on the $90^{\circ}$ model. The implication of these models is that SDSS\,J1205$-$0242 has a total system age between $2.5$ -- $10$\,Gyr.}
  \label{fig:sdss1205_bdmr}
  \end{center}
\end{figure}

In addition to the eclipse of the white dwarf, we also detect the transit of the white dwarf across the irradiated face of the brown dwarf, as shown in Fig.~\ref{fig:sececl}. The depth of this feature is strongly dependent upon the ratio of the radii and can be combined with the primary eclipse to place stringent constraints upon the inclination \citep[e.g.][]{parsons10,parsons12}. In this case, the $90^{\circ}$ model had a $\chi^2$ of 369 (fitting 217 points). Models with inclinations lower than $87^{\circ}$ had $\chi^2$ values higher than 434 ($\chi^2/\mathrm{DOF}>2$), since they predict eclipses that are too shallow. Additionally, the shape of the brown dwarf is more distorted in the lower inclination models (since it is closer to Roche lobe filling) leading to much poorer fits (e.g. the $85^{\circ}$ model in Fig.~\ref{fig:sececl}). Therefore, our secondary eclipse data place a lower limit of $87^{\circ}$ on the inclination.

Our final constraints on the stellar parameters are shown in Fig.~\ref{fig:sdss1205_wdmr} and Fig.~\ref{fig:sdss1205_bdmr} for the white dwarf and brown dwarf, respectively, and are listed in full in Table~\ref{tab:parameters}. We found that the minimum inclination of the system is $85^{\circ}$; below this the radius of the brown dwarf needs to be so large to fit the eclipse width that it fills its Roche lobe. This places a hard upper limit on the mass of the white dwarf of 0.43{\MSUN}. Our results from the secondary eclipse ($i > 87^{\circ}$) further constrain this upper limit to 0.40{\MSUN}. The uncertainty on the $K_1$ measurement dominates the error on the brown dwarf's mass, leading to a mass range of $0.049\pm0.06${\MSUN} ($51\pm6$\MJUP).

Fig.~\ref{fig:sdss1205_wdmr} shows that the measured radius of the white dwarf is fully consistent with theoretical predictions for helium-core white dwarfs (red line) by \citet{panei07} at an inclination of $90^{\circ}$ and is slightly oversized at lower inclinations, although this is still well within the uncertainties. 
The radii predicted by the carbon-oxygen core models (blue line) of \citet{fontaine01} are significantly smaller compared to our measurements. Both models have canonically thick hydrogen-layer masses, and we conclude that SDSS\,J1205$-$0242 has a helium core. At the highest inclinations the fit is also consistent with the white dwarf parameters found from the SOAR spectroscopy, implying that the true inclination is somewhere close to $90^{\circ}$. Fig.~\ref{fig:sdss1205_bdmr} shows that the brown dwarf's (volume-averaged) radius is consistent with theoretical predictions if it is older than 2.5\,Gyr ($>$3.5\,Gyr for the $90^{\circ}$ solution). We did not detect the brown dwarf in our $z'$ band light curves, placing an upper limit on its spectral type of L0, consistent with its classification as a brown dwarf.

We estimate the distance to the white dwarf by fitting the SDSS photometry with the \citet{panei07} models. We sample the posterior probability distributions for the parameter set $\{\log g,\, T_{\rm eff},\, {\rm E}(g-i),\, d\}$ using a Markov-Chain Monte-Carlo (MCMC) analysis. Posteriors on  $\{\log g,\, T_{\rm eff}\}$ come from the SOAR spectral fits, whilst the posterior on ${\rm E}(g-i)$ is uniform between 0 and the maximum extinction along the line of sight. To minimise the effects of contamination by the irradiated companion, we only fit the $u'g'r'$ photometry and find a distance of $720 \pm 40$ pc. Fitting the full $u'g'r'i'z'$ dataset does not change the distance significantly.

By combining this distance with an estimate of the proper motion based upon SDSS and PanSTARRS \citep{tian17} and adopting $\gamma_2$ as an estimate of the radial velocity of the system we can calculate its Galactic space velocity, relative to the local standard of rest, as $UVW = (36, -19, 35) \pm (7, 6, 4)$\,\kms. We adopt the convention that the sign of $U$ is positive towards the Galactic anti-centre. Following \cite{bensby14}, we find that SDSS\,J1205$-$0242 is ten times more likely to belong to the thin disk than the thick disk, and 50,000 times more likely to belong to the thin disk than the halo, justifying the adoption of solar metallicity models for the brown dwarf.

\section{A 72.5-min Binary: SDSS J1231+0041}

The second of our two systems, SDSS\,J1231+0041, is a faint ($g=20.1$) white dwarf identified from a serendipitous SDSS spectrum. Spectroscopic fits by \citet{rebassa16} to the SDSS spectrum found the white dwarf to be a $\TEFF=38740\pm2680$\,K $\log g = 7.07\pm0.41$ white dwarf; their solution suggested a possible photometric excess at longer wavelengths, sufficient for them to classify it as a possible white dwarf plus main-sequence star system.

We have fit our higher-S/N SOAR spectrum to better constrain the white dwarf atmospheric parameters, as we did for SDSS\,J1205$-$0242. Our updated SOAR fits find the primary white dwarf has $\TEFF=37210\pm1140$\,K, $\log{g}=7.77\pm0.15$, which yields a white dwarf mass of $0.56\pm0.07$\,{\MSUN} using the models of \citet{fontaine01}. The SOAR spectrum and best fit are shown in the right panel of Fig.~\ref{fig:soar_spec}. Following the same method as for SDSS\,J1205$-$0242, we estimate a distance of $1500 \pm 200$\,pc by fitting the carbon-oxygen core white dwarf models of \cite{fontaine01} to the SDSS $u'g'r'$ photometry.

\begin{figure}
  \begin{center}
    \includegraphics[width=\columnwidth]{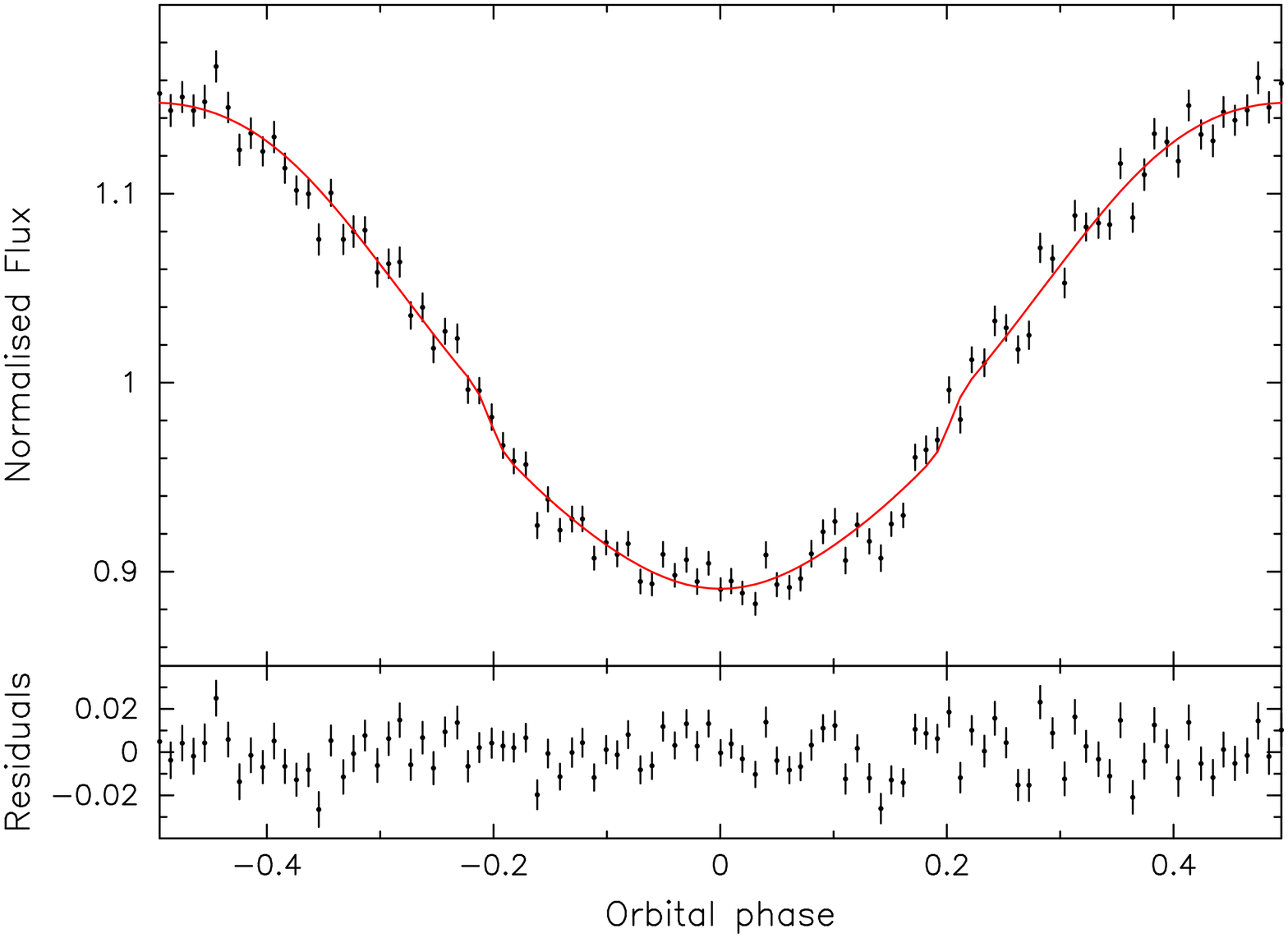}
    \includegraphics[width=\columnwidth]{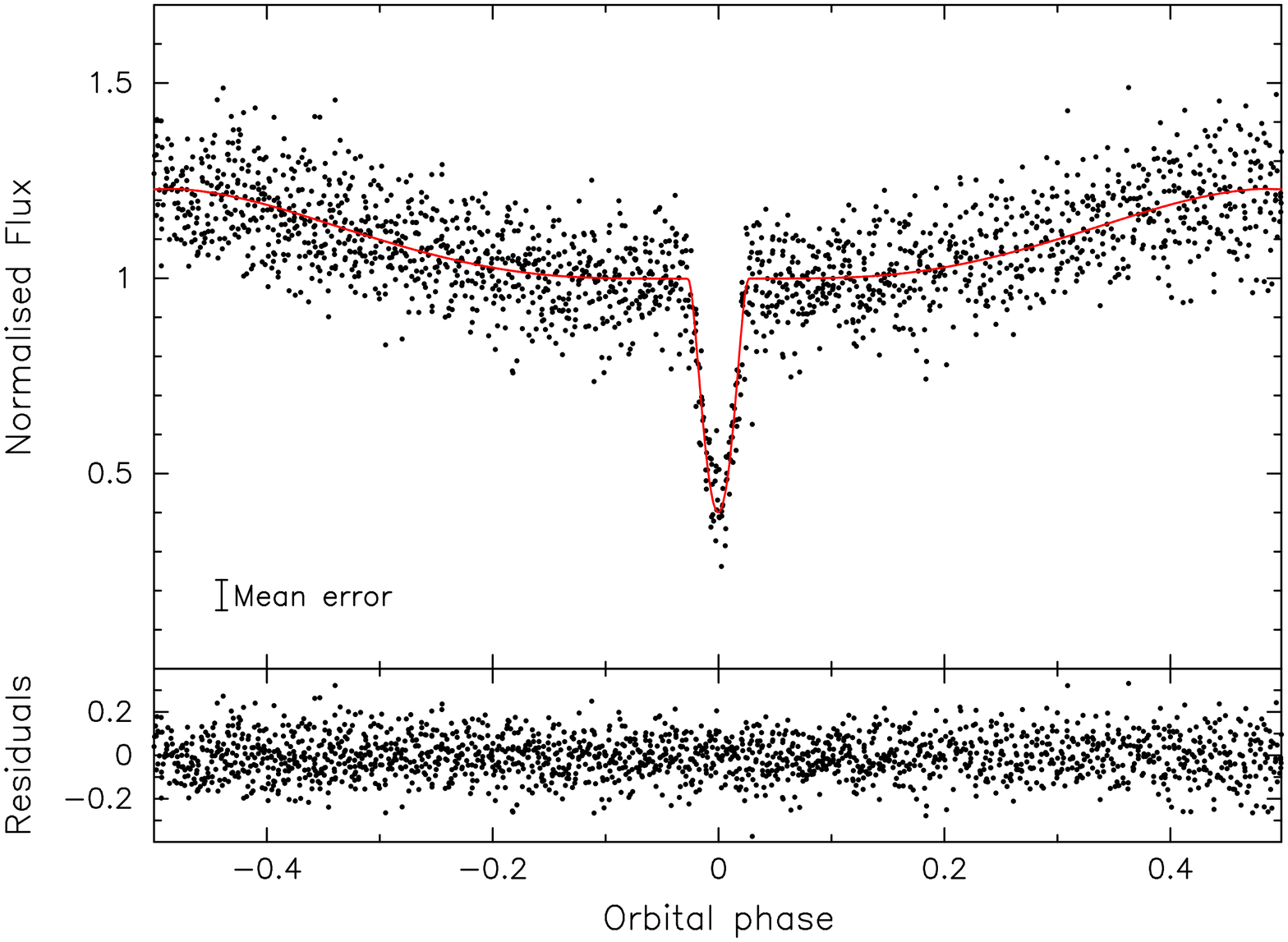}
    \vspace{-12pt}
    \caption{{\bf Top:} Phase-folded, binned {\em K2} light curve of SDSS\,J1231+0041. As with SDSS\,J1205$-$0242, the data are smeared since each long-cadence exposure comprises more than 40 per cent of the 72.5-min orbital period. Still, it shows a strong reflection effect, as well as slight evidence for an eclipse. {\bf Bottom:} Phase-folded, ULTRASPEC {\em kg5} light curve of SDSS\,J1231+0041 with higher time sampling. The partial eclipse of the white dwarf is clear, as is the out-of-eclipse reflection effect. A model light curve is over-plotted in red (and smeared in the top panel to match the {\em K2} exposures).} 
  \label{fig:sdss1231_lc}
  \end{center}
\end{figure}

The {\em K2} light curve of SDSS\,J1231+0041 shows clear variations on a period of 0.050353815(28) days (72.5 min), displayed in Fig.~\ref{fig:sdss1231_lc}. However, since these data were taken in long-cadence mode (with 29.4-min exposures) it was not immediately clear whether this is the true binary period, or if the period is twice this value. If this is the binary period, then the light curve variations must be the result of reprocessed light on the inner hemisphere of the companion to the white dwarf (i.e. reflection effect). The period could also be double this value, with the variations then caused by the Roche-distorted companion presenting different surface areas throughout the orbit (i.e. ellipsoidal modulation). The long exposure times of the {\em K2} data relative to the variability make it difficult to distinguish between these two possibilities. There is also marginal evidence for an eclipse in the form of a steeper curve just before and after the minimum.

Our ground-based follow-up high-speed photometry (Fig.~\ref{fig:sdss1231_lc}) shows that the true binary period is 72.5 min, and establishes that the system is eclipsing, albeit only partially (we do not detect any secondary eclipse). In total we covered four eclipses of the white dwarf (see Table~\ref{tab:etimes}). From these we determined the ephemeris for the system to be
\begin{eqnarray}
\mathrm{MJD(BTDB)} = 57746.435076(24) + 0.050353796(23)E.
\end{eqnarray}

\begin{figure}
  \begin{center}
    \includegraphics[width=\columnwidth]{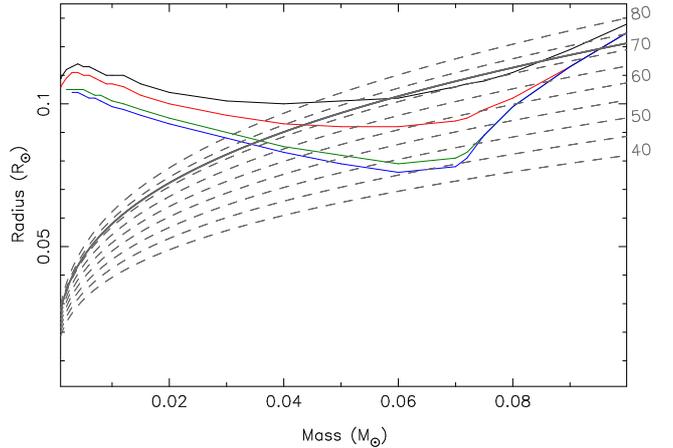}
    \vspace{-12pt}
    \caption{Theoretical mass--radius relationships (solid lines) for solar metallicity brown dwarfs and low-mass stars \citep{baraffe03} with ages of 0.5\,Gyr (black), 1\,Gyr (red), 5\,Gyr (green) and 10\,Gyr (blue). The grey dashed lines show the possible loci of Roche-lobe-filling companions to a white dwarf of mass $M_\mathrm{WD}=0.56$\MSUN\ for fixed orbital period (in steps of 5\,min). The figure shows that the most dense brown dwarfs are those with masses $\sim$0.065{\MSUN} that could still fit within their Roche lobes at periods as short as 45\,min, provided they are old enough. It also shows that, to fit within its Roche lobe at a period of 72.5\,min (indicated by the solid grey line), the companion to the white dwarf in SDSS\,J1231+0041 must have a mass of less than 0.095{\MSUN}.} 
  \label{fig:density}
  \end{center}
\end{figure}

Since SDSS\,J1231+0041 is only partially eclipsing, determining accurate parameters from the light curve is complicated due to the extra level of degeneracy. However, with such a short orbital period we can place an upper limit on the mass of the companion to the white dwarf, based on the fact that it does not fill its Roche lobe. In Fig.~\ref{fig:density} we show several mass-radius relationships for low-mass stars and brown dwarfs of different ages and solar metallicity. Also shown are lines of constant density at different orbital periods, which effectively shows the Roche lobe radius at different orbital periods. It is interesting to note how the radii of these low-mass objects are strongly related to their ages, to the extent that some binary configurations could only be possible with older brown dwarfs. For example, any detached system with a period $\lesssim$50 minutes must be older than $\sim$5\,Gyr and must have a mass $\lesssim$0.07{\MSUN}. With a period of 72.5 min we can only say that SDSS\,J1231+0041 must be older than $\sim$1\,Gyr, since at this period most of the models converge. We can also place an upper limit on the mass of the companion to the white dwarf of $\sim$0.095{\MSUN}; anything more massive than this would fill its Roche lobe. Therefore, it is quite likely that the companion in SDSS\,J1231+0041 is a brown dwarf, although radial-velocity data are required to confirm this.

Such a low-mass companion is completely outshone by the white dwarf at visible wavelengths, but \citet{rebassa16} suggested a possible photometric excess at red optical wavelengths. We have phased the SDSS photometry to the ephemeris established here and find it was all taken within 5 min of orbital phase 0.25. The apparent photometric red excess is therefore likely the result of irradiation, since the heated face of the companion has a much lower temperature but larger area than the white dwarf.

Our final parameters for both systems are listed in Table~\ref{tab:parameters}.
\begin{table}
 \centering
  \caption{Stellar and binary parameters for the two systems presented in this paper. WD refers to the white dwarf.}
  \label{tab:parameters}
  \begin{tabular}{@{}lcc@{}}
  \hline
  Parameter                  & SDSS\,J1205-0242 & SDSS\,J1231+0041 \\
  \hline
  Orbital period (d)         & 0.049465250(6)   & 0.050353796(23)  \\
  Orbital separation (\RSUN) & 0.42$-$0.45      & -                \\
  Orbital inclination (deg)  & 87$-$90          & -                \\
  WD mass (\MSUN)            & $0.39\pm0.02$    & $0.56\pm0.07$    \\
  WD radius (\RSUN)          & 0.0217$-$0.0223  & -                \\
  WD \TEFF (K)               & $23680\pm430$    & $37210\pm1140$   \\
  WD $\log{g}$               & $7.37\pm0.05$    & $7.77\pm0.14$    \\
  WD cooling age (Myr)       & 50               & 5                \\
  Secondary mass (\MSUN)     & $0.049\pm0.006$  & $\lesssim 0.095$ \\
  Secondary radius (\RSUN)   & 0.081$-$0.087    & $\lesssim 0.12$  \\
  App. magnitude ($g'$)      & 18.5             & 20.1             \\
  Distance (pc)              & $720\pm40$       & $1500\pm200$     \\
  \hline
\end{tabular}
\end{table}

\section{Discussion}

The white dwarfs in both the systems presented in this paper are the remnants of giant stars that were once much larger than their current orbits. This implies significant orbital shrinkage and points towards their emergence from common-envelopes that formed around both components of the binaries when mass transfer from the giant stars to their low mass companions took place \citep{1976IAUS...73...75P}. Common-envelope evolution is one of the most poorly understood and yet significant phases of close binary evolution \citep{2013A&ARv..21...59I}, and well-constrained examples of its effects are worth examination for the constraints they may raise.

In this case, SDSS\,J1205$-$0242 offers the most interesting test, first because it is better constrained, second because it contains a moderately low-mass, helium-core white dwarf. The helium white dwarf in SDSS\,J1205$-$0242 is a remnant of the first ascent red giant branch (RGB). As \cite{2000A&A...360.1011N} pointed out, the close relation between the core-mass and radius of RGB stars \citep{1970A&A.....6..426R} can allow tight constraints to be derived on the prior evolution of binary stars containing helium white dwarfs. The radius of an RGB star rises rapidly with the helium core mass \cite[approximately $\propto M_c^4$,][]{1985ApJS...58..661I}, thus helium white dwarfs of low mass that have emerged from common-envelopes are of particular interest since they come from relatively small, tightly-bound RGB stars. They can therefore lead to the most stringent constraints upon the efficiency with which the envelope is ejected. We express the effect of the common-envelope upon the orbital separation, $a$, through the relation 
\begin{equation}
\alpha \left( \frac{G M_{1f} M_2}{2 a_f} - \frac{G M_{1i} M_2}{2 a_i}\right)
= \frac{G M_{1i} (M_{1i} - M_{1f})}{\lambda R_{1i}}, \label{eq:ce}
\end{equation}
which equates a fraction of the orbital energy change on the left with the binding energy of the RGB's envelope on the right \citep{1984ApJ...277..355W,2000A&A...360.1043D}. Here the subscripts $i$ and $f$ refer to the initial and final values of the respective parameter when they differ. The parameters $\alpha$ and $\lambda$ encapsulate the efficiency with which orbital energy is used to eject the envelope and the internal structure of the envelope respectively.

There are alternative formulations for the binding energy of the envelope \citep{1993PASP..105.1373I}. We claim no advantage for our choice other than its popularity, which eases comparison with other studies; we refer the reader to  
\cite{2013A&ARv..21...59I} and \cite{2010A&A...520A..86Z} for further discussions of such variations and their effect upon the outcome of the common-envelope phase. Like
\cite{2010A&A...520A..86Z}, we condense what we can deduce from the system into a single constraint upon the combination parameter, $\alpha \lambda$.

\begin{figure}
  \begin{center}
    \includegraphics[width=\columnwidth]{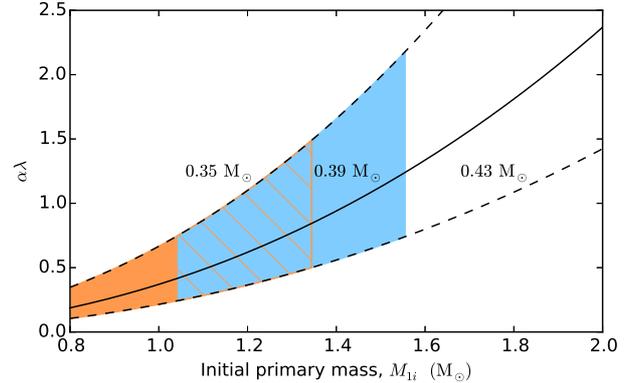}
    \vspace{-20pt}
\caption{
The value of the combined common-envelope / RGB structure parameter $\alpha
\lambda$ \protect\citep{2000A&A...360.1043D} required to match the parameters
of SDSS\,J1205$-$0242. The curved lines were calculated for three values of
white dwarf mass, the value of $0.39{\MSUN}$ from spectroscopy and $2\sigma$
either side of it and set $M_2 = 0.049\,$\MSUN. The shaded regions show the
ranges of progenitor mass consistent with the age of the brown dwarf for the
models of \protect\cite{baraffe15} (upper-right, shaded blue) and
\protect\cite{2008ApJ...689.1327S} (lower-left, shaded orange), allowing for
the white dwarf's cooling age of $50\,$Myr \protect\citep{panei07} and using
the formula for the time taken to reach the base of the RGB from
\protect\cite{2000MNRAS.315..543H}. Solar metallicity has been assumed
(Section~\protect\ref{sec:1205}).} 
  \label{fig:ce}
  \end{center}
\end{figure}

The core mass-radius relation means that $R_{1i}$ is largely defined by the mass of the white dwarf $M_{1f}$, with only a modest dependence upon its progenitor's mass $M_{1i}$. Therefore, as the progenitor mass increases, both terms in the numerator of the right-hand side of Eq.~\ref{eq:ce} increase with little change in the denominator. On the left-hand side, however, there is relatively little change with the progenitor mass, as it is the first term in the brackets that dominates, since $a_i \gg a_f$. The result is that the value of $\alpha \lambda$ required to produce SDSS\,J1205$-$0242 increases rapidly with the progenitor mass, $M_{1i}$. These constraints are encapsulated in Fig.~\ref{fig:ce}
which is based on the formulae presented by \cite{2000MNRAS.315..543H} and
\cite{1983ApJ...268..368E} in order to calculate $R_{1i}$ and $a_i$ for a given choice of progenitor mass.

Ranges of progenitor mass consistent with the age of the brown dwarf for two sets of models \citep{baraffe15,2008ApJ...689.1327S} are highlighted in Fig.~\ref{fig:ce}. The range of $\alpha \lambda$ runs from 0.1 -- 2.2, consistent with many of the systems studied in a similar manner by \cite{2010A&A...520A..86Z}. If brown dwarf models can be improved, there is potential for sharper constraints upon the common-envelope parameters, and, given the easily-detectable secondary eclipse in SDSS~J1205$-$0242 (Fig.~\ref{fig:sececl}), there are good prospects for tightening the parameter constraints significantly beyond those shown in Fig.~\ref{fig:ce}. There are caveats however: first are the significant existing uncertainties of cloud physics, molecular opacity and convection in brown dwarf models \citep{2008ApJ...689.1327S, baraffe15}, and second, it is possible that the unusual environment of rapid rotation and irradiation could affect the brown dwarf's size, although the 50~Myr since the common-envelope is a blink of an eye compared to the brown dwarf's Kelvin-Helmholtz timescale. Resolving such uncertainties is motivation for finding more examples of such systems.

The usual aim of studies such as this is to constrain the common-envelope efficiency parameter $\alpha$, but as we have seen it is the combination $\alpha \lambda$ that is directly constrained. Unfortunately, the structure parameter $\lambda$, often taken to be $0.5$ \citep{1990ApJ...358..189D}, is almost as ill-defined as $\alpha$, as it is not known to what extent the internal (thermal) energy of the envelope needs to be taken into account when calculating it \citep{1995MNRAS.272..800H,2000A&A...360.1043D,2014A&A...566A..86C}; a contribution from internal energy can increase $\lambda$ significantly. The relatively tightly-bound RGB star in this instance should make this uncertainty relatively small compared to later stages of stellar evolution, and once the parameters of the binary are firmed up, it will be worth investigating stellar models to define the range of $\lambda$ for this specific instance. In any event, it is clear that SDSS\,J1205$-$0242 and similar white dwarf / brown dwarf systems have significant potential for both common-envelope evolution and brown dwarf physics.

Prior to the formation of the common-envelope, the binary would have had an orbital period in the range $60-200$\,days, placing it within or close to the ``brown dwarf desert'' where few brown dwarf companions to solar-type stars are seen \citep{2000PASP..112..137M,2014MNRAS.439.2781M}. It would be interesting to ascertain whether or not the numbers of white dwarf / brown dwarf PCEBs are consistent with the rarity of their progenitors in radial-velocity surveys.

These systems will transfer mass in the near future ($\sim$300\,Myr for SDSS\,J1205$-$0242), and presumably appear as cataclysmic variable stars. Their existence in this form, however, may be brief, if recent suggestions of the destabilising effects of novae on cataclysmic variables containing low mass white dwarfs are correct \citep{2016MNRAS.455L..16S,2016ApJ...817...69N}. They may then soon merge to become single white dwarfs, and, in the case of SDSS\,J1205$-$0242, a single white dwarf of low mass \citep{2017MNRAS.466L..63Z}, a number of which are known \citep{marsh95,2011ApJ...730...67B}. Their emergence from the common-envelope phase so close to Roche-lobe filling also suggests that had the companions been even less massive, these systems might not have survived the common-envelope at all but simply have merged. This exact scenario has been suggested as another way to form single low-mass white dwarfs \citep{1998A&A...335L..85N}.

\section{Conclusions}

Using long-cadence photometry from the {\em Kepler} space telescope we have discovered two new ultrashort, detached eclipsing binaries composed of white dwarfs plus cool companions. The binaries have such short orbital periods --- 71.2\,min and 72.5\,min --- that the companions are likely substellar on the basis of their periods alone, in order that they do not fill their respective Roche lobes.

Follow-up photometry and spectroscopy significantly constrain both systems. SDSS\,J1205$-$0242 contains a hot ($\TEFF=23680\pm430$\,K), low-mass ($0.39\pm0.02${\MSUN}) white dwarf with a radius consistent with a helium-core white dwarf (0.0217$-$0.0223{\RSUN}). It is totally eclipsed every 71.2\,min by a $45$--$57\,${\MJUP} brown dwarf companion that has a radius consistent with an age greater than 2.5\,Gyr (0.081$-$0.087{\RSUN}). Detection of secondary eclipses constrains the orbital inclination to $>87$\,deg.

SDSS\,J1231+0041 contains a hot ($\TEFF=37210\pm1140$\,K), $0.56\pm0.07${\MSUN} white dwarf that is partially eclipsed every 72.5\,min by a companion of less than 0.095{\MSUN}, likely also to be a brown dwarf. Details of all physical constraints to both systems are listed in Table~\ref{tab:parameters}.

The shorter-period system, SDSS\,J1205$-$0242, places useful constraints upon common-envelope evolution, because of its helium-core white dwarf and the need for the white dwarf's total age to match the age of its brown dwarf companion. This demonstrates that ultrashort-period white dwarf plus brown dwarf binaries can be used to test theories of common-envelope evolution, because of the time-dependent radii of brown dwarfs, although uncertainties in brown dwarf models require clarification for this method to be applied with confidence.

Both systems were discovered as part of a search for transits and variability among white dwarfs in {\em K2} Campaign 10; we expect to find more similar short-period eclipsing binaries as {\em K2} continues surveying new fields along the ecliptic. The results here also help build expectations for the next space-based photometric mission, {\em TESS} \citep{ricker14}, which can be used to target many bright white dwarfs all-sky at 2-min cadence.

\section*{Acknowledgements}

SGP acknowledges the support of the Leverhulme Trust. The research leading to these results has received funding from the European Research Council under the European Union's Seventh Framework Programme (FP/2007-2013) / ERC Grant Agreement numbers 340040 (HiPERCAM) and 320964 (WDTracer), as well as the European Union's Horizon 2020 Research and Innovation Programme / ERC Grant Agreement number 677706 (WD3D). ULTRACAM, TRM, VSD, SPL are supported by the Science and Technology Facilities Council (STFC) under grants ST/L000733 and ST/M001350. DAHB is supported by the National Research Foundation of South Africa. Support for this work was provided by NASA through Hubble Fellowship grant \#HST-HF2-51357.001-A, by NASA {\em K2} Cycle 4 Grant NNX17AE92G, as well as NSF grants AST-1413001 and AST-1312983. This work has made use of data obtained at the Thai National Observatory on Doi Inthanon, operated by NARIT; the Southern Astrophysical Research (SOAR) telescope, which is a joint project of the Minist\'{e}rio da Ci\^{e}ncia, Tecnologia, e Inova\c{c}\~{a}o da Rep\'{u}blica Federativa do Brasil, the U.S. National Optical Astronomy Observatory, the University of North Carolina at Chapel Hill, and Michigan State University; the McDonald Observatory of the University of Texas at Austin; as well as the Southern African Large Telescope (SALT), through DDT programme 2016-2-DDT-006, where the assistance of Marissa Kotze is acknowledged. Data for this paper have been obtained under the International Time Programme of the CCI (International Scientific Committee of the Observatorios de Canarias of the IAC) with the Gran Telescopio Canarias (GTC) operated on the island of La Palma in the Observatorio del Teide/Roque de los Muchachos.

{\it Facilities:} {\em K2}, Otto Struve (ProEM), TNT (ULTRASPEC), SOAR (Goodman), SALT (SALTICAM), GTC (OSIRIS)


\bibliographystyle{mnras}
\bibliography{k2_wdbd}


\bsp	
\label{lastpage}
\end{document}